\documentclass[prd,aps,amsmath,nofootinbib]{revtex4}
\usepackage{graphicx}
\usepackage{bm}
\usepackage{epsfig}

  \newcount\hour \newcount\minute
  \hour=\time \divide \hour by 60
  \minute=\time
  \newcommand{\mydate}{\ \today \ - \number\hour :\ifnum \minute<10 0\fi 
\number\minute}

\def\Dslash{D\!\!\!\!\slash}

\def\bnslash{\bar n\!\!\!\slash}

\def\OMIT#1{}

\newcommand{\nn}{\nonumber} 

\newcommand{\bn}{{\bar n}}
\newcommand{\bea}{\begin{eqnarray}}
\newcommand{\eea}{\end{eqnarray}}

\newcommand{\bnP}{\bar {\cal P}}

\newcommand{\mcdot}{\!\cdot\!}

\newcommand{\bfk}{{\bf k}}
\newcommand{\SCETa}{\mbox{${\rm SCET}_{\rm I}$ }}
\newcommand{\SCETb}{\mbox{${\rm SCET}_{\rm II}$ }}

\def\lqcd{\Lambda_{\rm QCD}}

\begin{document}


\preprint{ \vbox{\hbox{hep-ph/0312302}  }}

\title{Infrared regulators and \SCETb}

\author{Christian W.~Bauer}
\email{bauer@theory.caltech.edu}
\affiliation{California Institute of Technology, Pasadena, CA 91125}
\author{Matthew P.~Dorsten}
\email{dorsten@theory.caltech.edu}
\affiliation{California Institute of Technology, Pasadena, CA 91125}
\author{Michael P.~Salem}
\email{salem@theory.caltech.edu}
\affiliation{California Institute of Technology, Pasadena, CA 91125}


\begin{abstract}
We consider matching from \SCETa, which includes ultrasoft and
collinear particles, onto \SCETb with soft and collinear particles at
one loop. Keeping the external fermions off their mass shell does not
regulate all IR divergences in both theories. We give a new
prescription to regulate infrared divergences in SCET. Using this
regulator, we show that soft and collinear modes in \SCETb are 
sufficient to reproduce all the infrared divergences of \SCETa. We explain
the relationship between IR regulators and an additional mode proposed for \SCETb.

\end{abstract}

\maketitle

\section{Introduction}

Soft-collinear effective theory~\cite{BFL,BFPS,BS,BPS} describes the
interactions of soft and ultrasoft (usoft) particles with collinear
particles. Using light-cone coordinates in which a general
four-momentum is written as \mbox{$p^\mu = (p^+,p^-,p^\perp) = (n \!\cdot\!
p,\bn\!\cdot\! p,p^\perp)$}, where $n$ and $\bn$ are four-vectors on the
light cone ($n^2=\bn^2=0$, $n \cdot \bn = 2$), these three degrees of
freedom are distinguished by the scaling of their momenta:
\begin{eqnarray}
\begin{tabular}{ll}
\mbox{collinear:} & $p_c^\mu \sim Q(\lambda^2, 1, \lambda)$, \\
\mbox{soft:} & $p_s^\mu \sim Q(\lambda,\lambda,\lambda)$, \\
\mbox{usoft:} &$ p_{us}^\mu \sim Q(\lambda^2,\lambda^2,\lambda^2)$.
\end{tabular}
\end{eqnarray}
The size of the expansion parameter $\lambda$ is determined by the
typical off-shellness of the collinear particles in a given
problem. For example, in inclusive decays one typically has $p_c^2
\sim Q^2 \lambda^2 \sim Q\lqcd$, from which it follows that $\lambda =
\sqrt{\lqcd/Q}$. For exclusive decays, however, one needs collinear
particles with $p_c^2 \sim \lqcd^2$, giving $\lambda =
\lqcd/Q$. One is usually interested in describing the interactions
of these collinear degrees of freedom with non-perturbative degrees of
freedom at rest, which satisfy $p^\mu \sim (\lqcd,\lqcd,\lqcd)$. Thus
inclusive processes involve interactions of collinear
and usoft degrees of freedom, while exclusive decays are described by 
interactions of collinear and soft degrees of freedom.
The
theory describing the former set of degrees of freedom is called
\SCETa, while the latter theory is called \SCETb \cite{ffprl}.

Interactions between usoft and collinear degrees of freedom are 
contained in the leading-order Lagrangian of \SCETa,
\begin{eqnarray}
{\cal L_{\rm I}} = \bar \xi_n \left[ i n \!\cdot\! D + i \Dslash_c^\perp 
\frac{1}{i \bn\!\cdot\! D_c}i \Dslash_c^\perp  \right] \frac{\bnslash}{2} \xi_n\,,
\end{eqnarray}
and are well understood. The only interaction between collinear fermions and usoft gluons is from the derivative
\begin{eqnarray}
iD^\mu = iD_c^\mu + g A_{us}^\mu\,.
\end{eqnarray}
These interactions can be removed from the
leading-order Lagrangian by the field redefinition \cite{BPS}
\begin{eqnarray}\label{Yredef}
 \xi_n = Y_n \xi_n^{(0)}\,, \quad 
 A_n = Y_n A_n^{(0)} Y_n^\dagger \, , \qquad \nn\\
 Y_n(x) = \mbox{P}\exp\left( ig \! \int_{-\infty}^0 \!\!\!\!\!\!\! ds\: n\mcdot 
 A_{us}(x+ns) \right)\,.
\end{eqnarray}
However, the same field redefinition has to be performed on the
external operators in a given problem, and this reproduces the
interactions with the usoft degrees of freedom. Consider for example
the heavy-light current, which in \SCETa is given by
\begin{eqnarray}\label{heavylightI}
J_{hl}^I(\omega) = \left[\bar \xi_n W_n\right]_\omega \Gamma h_v\, ,
\end{eqnarray}
where $h_v$ is the standard field of heavy quark effective
theory~\cite{HQET}, the Wilson line $W_n$ is required to ensure
collinear gauge invariance~\cite{BS} and $\omega$ is the large momentum label of the gauge invariant $[\bar \xi_n W_n]$ collinear system. Written in terms of the
redefined fields, this current is
\begin{eqnarray}\label{heavylightIY}
J_{hl}^{I}(\omega) = \left[\bar \xi_n^{(0)} W_n^{(0)}\right]_\omega 
\Gamma \left[Y^\dagger_n h_v\right]\, .
\end{eqnarray}

For exclusive decays, we need to describe the interactions of soft
with collinear particles. This theory is called \SCETb \cite{ffprl}. Since adding a
soft momentum to a collinear particle takes this particle off its mass
shell $(p_c + p_s)^2 \sim (Q\lambda,Q,Q\lambda)^2 \sim Q^2 \lambda
\sim Q \lqcd$, there are no couplings of soft to collinear
particles in the leading-order Lagrangian.\footnote{At higher orders,
higher dimensional operators with at least two soft and two collinear
particles can appear.} Thus, the Lagrangian is given by \cite{NeubertSCET2,messenger}
\begin{eqnarray}
{\cal L_{\rm II}} = \bar \xi_n \left[ i n \!\cdot\! D_c + i
\Dslash_c^\perp \frac{1}{i \bn\!\cdot\! D_c}i \Dslash_c^\perp  \right]
\frac{\bnslash}{2} \xi_n\, .
\end{eqnarray}
In this theory, the heavy-light current is given by
\begin{eqnarray}
\label{heavylightII}
J_{hl}^{II}(\omega, \kappa) = \left[\bar \xi_n^{(0)} W_n^{(0)}\right]_\omega 
\Gamma \left[S^\dagger_n h_v\right]_\kappa \, ,
\end{eqnarray}
where $S_n$ is a soft Wilson line in the $n$ direction defined by
\begin{eqnarray}
 S_n(x) = \mbox{P}\exp\left( ig \! \int_{-\infty}^0 \!\!\!\!\!\!\! ds\: n\mcdot 
 A_{s}(x+ns) \right)\,.
\end{eqnarray}
This is the most general current invariant under collinear and soft
gauge transformations.

This paper is organized as follows: We first consider matching the heavy-light current in \SCETa onto the heavy-light current in \SCETb using off-shell fermions. While the terms logarithmic in the off-shellness do not agree in the two theories, we argue that this is due to unregulated IR divergences in \SCETb. We then discuss IR regulators in SCET in more detail. We first identify the problems with SCET regulators and then propose a new regulator that addresses these issues. Using this regulator we then show that soft and collinear modes in \SCETb are sufficient to reproduce the IR divergences of \SCETa and explain the relationship between IR regulators and an additional mode proposed for  \SCETb\cite{messenger}.

\section{Matching from \SCETa onto \SCETb}
\label{sec:match}

The only difference between \SCETa and \SCETb is the typical off-shellness
of the collinear degrees of freedom in the theory. The theory \SCETa
allows fluctuations around the classical momentum with $p_c^2 \sim
Q\lqcd$, while the theory \SCETb allows fluctuations with only $p_c^2
\sim \lqcd^2$. Since both theories expand around the same limit,
\SCETb can be viewed as a low energy effective theory of \SCETa.
Therefore, one can match from the theory \SCETa onto \SCETb by
integrating out the ${\cal O}(\sqrt{Q\lqcd})$ fluctuations.
 
To illustrate this matching, we consider the heavy-light
current. Using the definitions of this current given in
Eqs.~(\ref{heavylightI}) and (\ref{heavylightII}), we can write
\begin{eqnarray}
J_{hl}^I(\omega) = \int\!\! d\kappa \,\, C(\omega,\kappa)
\,\,J_{hl}^{II}(\omega,\kappa) \, .
\end{eqnarray}
At tree level one finds trivially
\begin{eqnarray}
C(\omega,\kappa) = 1\, .
\end{eqnarray}
In fact, this Wilson coefficient remains unity to all orders in perturbation theory, as was argued in Ref.~\cite{scetgauge}.

\begin{figure}[t]
\begin{center}
\includegraphics[bb=350 715 585 760, clip, width=.8\textwidth]{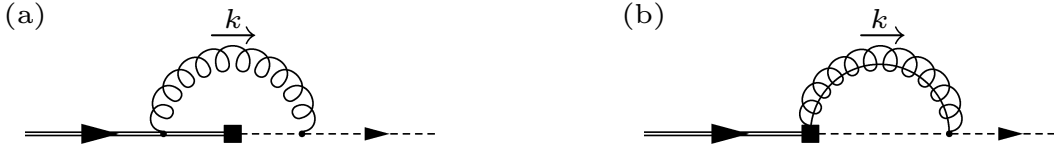}
\caption{Diagrams in \SCETa contributing to the matching. The solid
square denotes an insertion of the heavy-light current.}
\label{fig1}
\end{center}
\end{figure}

\begin{figure}[t]
\begin{center}
\includegraphics[bb=350 715 585 760, clip, width=.8\textwidth]{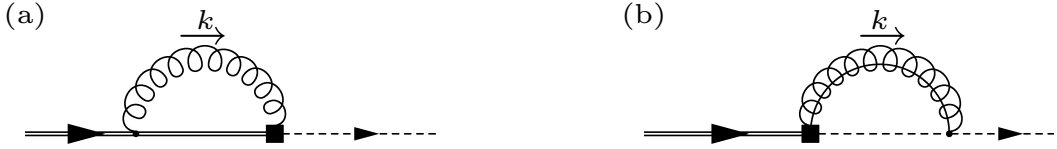}
\caption{Diagrams in \SCETb contributing to the matching.}
\label{fig2}
\end{center}
\end{figure}

To determine the matching coefficient at one loop, we calculate matrix
elements of the current in the two theories. There are two diagrams in
\SCETa, shown in Fig.~\ref{fig1}. For on-shell external states, we
find for the two integrals
\begin{eqnarray}
iA^{Ia} &=& g^2 C_F \mu^{4-d}\! \int \!\! \frac{{\rm d}^dk}{(2\pi)^d} 
\frac{1}{[-n \!\cdot\! k+i0][-v \!\cdot\! k+i0][k^2+i0]}\, , \\
iA^{Ib} &=& 2 g^2 C_F \mu^{4-d}\! \int \!\! \frac{{\rm d}^dk}{(2\pi)^d} 
\frac{\bn \!\cdot\! (p_c-k)}{[-\bn \!\cdot\! k+i0]
[k^2-2p_c \!\cdot\! k+i0][k^2+i0]}\, .
\end{eqnarray}
The diagrams in \SCETb are shown in Fig.~\ref{fig2}. For on-shell 
external states the two integrals are exactly the same as in \SCETa:
\begin{eqnarray}
iA^{IIa} &=& g^2 C_F \mu^{4-d}\! \int \!\! \frac{{\rm d}^dk}{(2\pi)^d} 
\frac{1}{[-n \!\cdot\! k+i0][-v \!\cdot\! k+i0][k^2+i0]}\, , \\
iA^{IIb} &=& 2 g^2 C_F \mu^{4-d}\! \int \!\! \frac{{\rm d}^dk}{(2\pi)^d} 
\frac{\bn \!\cdot\! (p_c-k)}{[-\bn \!\cdot\! k+i0]
[k^2-2p_c \!\cdot\! k+i0][k^2+i0]}\, .
\end{eqnarray}
Since the integrands are exactly the same, the loop diagrams will 
precisely cancel in the matching calculation. Thus we find that the 
Wilson coefficient $C(\omega,\kappa)$ remains unity, even at one loop. This confirms the arguments in \cite{scetgauge} to this order. 

The fact that both of these integrals are scaleless and therefore zero
might bother some readers. The vanishing of these diagrams is due to
the cancellation of collinear, infrared (IR) and ultraviolet (UV)
divergences. Introducing an IR regulator will separate these
divergences, and the UV will be regulated by dimensional
regularization. In Ref.~\cite{BFL} a small off-shellness was
introduced to regulate the IR divergences of \SCETa. In
Refs.~\cite{messenger} the divergence structure of \SCETb
was studied keeping both the heavy and the collinear fermions
off-shell. Using this IR regulator, the authors of
Refs.~\cite{messenger} argued that \SCETb does not
reproduce the IR divergences of \SCETa and introduced a new mode in
\SCETb to fix this problem. To gain more insight into their argument,
we will go through their calculation in some detail.

In \SCETa the first diagram is
\begin{eqnarray}
A^{Ia}_{p_c} &=& -ig^2 C_F \mu^{4-d}\! \int \!\! \frac{{\rm d}^dk}{(2\pi)^d} \frac{1}
{[\tilde p_c-n \!\cdot\! k+i 0][v \!\cdot\! (p_s-k)+i 0][k^2+i 0]} 
\nonumber\\
&=& -\frac{g^2 C_F}{2 \pi} (4 \pi)^{1-d/2} \Gamma\left(2-\frac{d}{2}\right) 
\mu^{4-d}\! \int^{\infty}_0 \!\!\!\!\!\! {\rm d}n \!\cdot\! k 
\left(n \!\cdot\! k - \tilde p_c\right)^{-1} n \!\cdot\! k^{d/2-2} 
\left( n \!\cdot\! k - 2 v \!\cdot\! p_s \right)^{d/2-2}
\nonumber\\
&=& \frac{\alpha_s C_F}{4 \pi} \left[
-\frac{1}{\epsilon^2} 
+ \frac{2}{\epsilon}  \log \frac{-\tilde p_c}{\mu}-2\log^2 \frac{-\tilde p_c}{\mu} 
+ 2\log \left(1-\frac{2v\!\cdot\! p_s}{\tilde p_c}\right)  \log \frac{2v\!\cdot\! p_s}{\tilde p_c}+ 2 \, {\rm Li}_2\left(\frac{2v\!\cdot\! p_s}{\tilde p_c}\right) - \frac{3 \pi^2}{4}
 \right]\, ,
\end{eqnarray}
where $d=4-2\epsilon$ and 
\begin{eqnarray}
\tilde p_c = \frac{p_c^2}{\bn \!\cdot\! p_c}\, .
\end{eqnarray}
In going from
the first line to the second, we closed the $\bn \!\cdot\! k$ contour
below, thus restricting $n \!\cdot\! k$ to positive values, and performed the
Euclidean $k_\perp$ integral.  The second
diagram gives
\begin{eqnarray}
\label{AIIbresult}
A^{Ib}_{p_c} &=& -2 i g^2 C_F \mu^{4-d} \! \int \!\! \frac{{\rm d}^dk}{(2\pi)^d} 
\frac{\bn \!\cdot\! (p_c-k)}{[-\bn \!\cdot\! k+i0]
[(k-p_c)^2+i0][k^2+i0]}
\nonumber\\
&=& \frac{\alpha_s C_F}{4 \pi} \left[ \frac{2}{\epsilon^2} + \frac{2}{\epsilon}- \frac{2}{\epsilon} \log \frac{-p_c^2}{\mu^2} 
+ \log^2 \frac{-p_c^2}{\mu^2} - 2 \log \frac{-p_c^2}{\mu^2} + 4 - \frac{\pi^2}{6}
\right]\, .
\end{eqnarray}
In this diagram it is necessary to choose $d < 4$ for the $k_\perp$ integral, but one requires $d > 4$ for the $\bn \!\cdot\! k$ integral. In the former integral, dimensional regularization regulates the divergence at $k_\perp = \infty$, while in the  latter it regulates the divergence at 
$\bn\!\cdot\! k=0$.  Both of these divergences have to be interpreted as UV, as discussed in section~\ref{sec:IRregulators}. Each diagrams contain a mixed UV-IR divergence of the form
$\log p_c^2/\epsilon$. This mixed divergence cancels in the sum of
the two diagrams and we find, after also adding the wave function
contributions,
\begin{eqnarray}
\label{A1psq}
A^I_{p_c} &=& \frac{\alpha_s C_F}{4 \pi} \left[ \frac{1}{\epsilon^2} +
\frac{2}{\epsilon} \log \frac{\mu}{\bn \!\cdot\! p_c}  + \frac{5}{2\epsilon}
+ \log^2 \frac{-p_c^2}{\mu^2}  - 2 \log^2  \frac{-\tilde p_c}{\mu}
-\frac{3}{2} \log \frac{-p_c^2}{\mu^2}- 2 \log  \frac{-2 v \!\cdot\! p_s}{\mu}\right.\nonumber\\
&&\qquad \left.
+ 2\log \left(1-\frac{2v\!\cdot\! p_s}{\tilde p_c}\right)  \log \frac{2v\!\cdot\! p_s}{\tilde p_c}+ 2 {\rm Li}_2\left(\frac{2v\!\cdot\! p_s}{\tilde p_c}\right) 
+\frac{11}{2} - \frac{11 \pi^2}{12}\right]\, .
\end{eqnarray}
This reproduces the IR behavior of full QCD. 

\begin{figure}[b]
\begin{center}
\includegraphics[bb=490 715 585 760, clip, width=.33\textwidth]{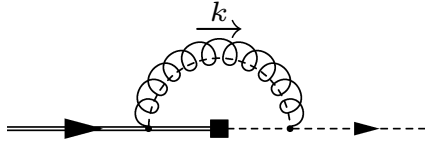}
\caption{Contribution of the additional \SCETb mode proposed in
Refs.~\cite{messenger}.}
\label{fig2c}
\end{center}
\end{figure}

Now consider the \SCETb diagrams. The second is identical to the one
in \SCETa:
\begin{eqnarray}
A^{IIb}_{p_c} = A^{Ib}_{p_c}\, .
\end{eqnarray}
The first diagram, however, is different and we find
\begin{eqnarray}
\label{AIIadetails}
A^{IIa}_{p_c} &=& -ig^2 C_F \mu^{4-d}\! \int \!\! \frac{{\rm d}^dk}{(2\pi)^d}  
\frac{1}{[-n \!\cdot\! k+i0] [v \!\cdot\! (p_s-k)+i0] [k^2+i0]} 
\nonumber\\
&=& -\frac{g^2 C_F}{2 \pi} \mu^{4-d} \int^{\infty}_0 \!\!\!\!\!\! 
{\rm d} n \!\cdot\! k \! \int \!\! \frac{{\rm d}^{d-2}k_\perp}{(2\pi)^{d-2}} 
\frac{1}{n \!\cdot\! k \, (k_\perp^2 + n \!\cdot\! k^2- 2 v \!\cdot\! p_s \, 
n \!\cdot\! k)}
\nonumber\\
&=& -\frac{\alpha_s C_F}{2\pi} (4 \pi)^{2-d/2} \Gamma\left(2-\frac{d}{2}\right) 
\mu^{4-d} \! \int^{\infty}_0 \!\!\!\!\!\! {\rm d} n \!\cdot\! k \, n \!\cdot\! k^{d/2-3} 
(n \!\cdot\! k - 2 v \!\cdot\!
p_s)^{d/2-2}\, .
\end{eqnarray}
Note that convergence of this integral at $n \!\cdot\! k = \infty$
requires $d < 4$, whereas convergence at $n \!\cdot\! k = 0$ requires
$d > 4$. In this case dimensional regularization is regulating both a
UV divergence at $n \!\cdot\! k = \infty$, as well as the divergence
at $n \!\cdot\! k = 0$ , which is IR in
nature, as we will discuss in section~\ref{sec:IRregulators}. Using the variable transformation $x=n\!\cdot\! k/(n\!\cdot\! k - 2v\!\cdot\! p_s)$ to relate this integral to a beta
function~\cite{Hillprivate} one finds
\begin{eqnarray}
\label{AIIaneubertresult}
A^{IIa}_{p_c} &=& \frac{\alpha_s C_F}{4\pi} \left[ \frac{1}{\epsilon^2}-
\frac{2}{\epsilon}\log \frac{-2 v \!\cdot\! p_s}{\mu} + 2 \log^2 \frac{-2 v
\!\cdot\! p_s}{\mu} + \frac{5 \pi^2}{12} \right]\, .
\end{eqnarray}
Adding the two diagrams together with the wave function contributions gives
\begin{eqnarray}
\label{A2psqinterm}
\frac{\alpha_s C_F}{4\pi} \left[ \frac{3}{\epsilon^2} -
\frac{2}{\epsilon}\log \frac{2 v \!\cdot\! p_s\,  p_c^2}{\mu^3} +
\frac{5}{2\epsilon} + \log^2 \frac{-p_c^2}{\mu^2} + 2 \log^2 \frac{-2 v
\!\cdot\! p_s}{\mu} - \frac{3}{2} \log \frac{-p_c^2}{\mu^2}- 2 \log  \frac{-2
v \!\cdot\! p_s}{\mu} 
+ \frac{11}{2} + \frac{\pi^2}{4}
\right]\, .
\end{eqnarray}
We can see that in the sum of the two diagrams the terms proportional
to $\log p_c^2 / \epsilon$ or $\log v\!\cdot\! p_s / \epsilon$ do not cancel
as they did in \SCETa. Furthermore, the finite terms logarithmic in
$p_c^2$ or $v \!\cdot\! p_s$ do not agree with the corresponding terms in
the \SCETa result. This fact prompted the authors of
Refs.~\cite{messenger} to conclude that \SCETb does not
reproduce the IR divergences of \SCETb and that a new mode is needed
in the latter effective theory. However, as we mentioned above, there are
problems with IR effects in this diagram. In fact, as we will show in
great detail in the next section, the off-shellness of the fermions
does not regulate all IR divergences in this diagram. This means that
the fact that the terms logarithmic in the fermion off-shellness do
not agree between \SCETa and \SCETb does not imply that the IR
divergences are not reproduced correctly since some $1/\epsilon$ poles
are IR in origin.

We also calculate the diagram in \SCETb containing
the additional mode proposed in
Refs.~\cite{messenger}. The new messenger mode has momenta
scaling as $p_{sc}^\mu \sim (\lqcd^2/Q, \lqcd,\lqcd^{3/2}/Q^{1/2})$. (Note
that the invariant mass of this term satisfies $p_{sc}^2 \ll
\lqcd^2$.) The diagram is shown in Fig.~\ref{fig2c} and we find
\begin{eqnarray}
A^{IIc}_{p_c} &=& -2 ig^2 C_F \mu^{4-d}\! \int \!\! \frac{{\rm d}^dk}{(2\pi)^d}  
\frac{1}{[\tilde p_c-n \!\cdot\! k+i 0][2 v \!\cdot\! p_s-\bn  \!\cdot\! k+i 0]
[k^2+i 0]} \nonumber\\
&=&  \frac{\alpha_s C_F}{4\pi} \left[ -\frac{2}{\epsilon^2} +
\frac{2}{\epsilon} \log \frac{2 v \!\cdot\! p_s \, \tilde p_c}{\mu^2} - \log^2  \frac{2 v \!\cdot\! p_s \, \tilde p_c}{\mu^2}
- \frac{\pi^2}{2}
 \right]\, .
\end{eqnarray}
Adding this term to the Eq.~(\ref{A2psqinterm}) cancels the 
terms proportional to $\log (2v \!\cdot\! p_s \, p_c^2/\mu^3)/\epsilon$ and  we find
\begin{eqnarray}
\label{A2psq}
A^{II}_{p_c} &=& \frac{\alpha_s C_F}{4 \pi} \left[ 
\frac{1}{\epsilon^2} +
\frac{2}{\epsilon} \log \frac{\mu}{\bn \!\cdot\! p_c}  + \frac{5}{2\epsilon}
+ \log^2 \frac{-p_c^2}{\mu^2} 
-\frac{3}{2} \log \frac{-p_c^2}{\mu^2}- 2 \log  \frac{-2 v \!\cdot\! p_s}{\mu}
\right.
\nonumber\\
&&\qquad \left.
+2\log^2 \left(\frac{-2 v \!\cdot p_s}{\mu}\right) - \log^2\left( \frac{2 v\!\cdot p_s \tilde p_c}{\mu^2}\right)
+ \frac{11}{2} - \frac{\pi^2}{4}\right]\, .\
\end{eqnarray}
This result does not agree with the \SCETa expression in Eq.~(\ref{A1psq}). However, this is expected, since the off-shellness in \SCETb satisfies $\tilde p_c \ll v\!\cdot\! p_s$. In this limit the \SCETa result in Eq.~(\ref{A1psq}) agrees with the result in Eq.~(\ref{A2psq}). 

\section{Infrared regulators in SCET}
\label{sec:IRregulators}

\subsection{Problems with known IR regulators}

One of the most important properties of \SCETa is the field
redefinition given in Eq.~(\ref{Yredef}), which decouples the usoft
from the collinear fermions. It is the crucial ingredient for proving
factorization theorems. Furthermore, only after performing this field
redefinition is it simple to match from \SCETa onto \SCETb, since one
can identify the Wilson line $Y_n$ in \SCETa with the Wilson line
$S_n$ in \SCETb. However, it is a well known fact that field
redefinitions only leave on-shell Green functions
invariant~\cite{offshellGreen}. Hence, the off-shellness of the collinear quark $p_c^2$ used to
regulate the IR in \SCETa takes away our ability to perform this field
redefinition. Since no field redefinition is performed on the heavy quark, one is free to give it an off-shellness.

IR divergences appear in individual diagrams, but they cancel in the set of diagrams contributing to a physical observable. More specifically, the IR divergences in virtual loop diagrams are cancelled against those in real emissions, which physically have to be included due to finite detector resolutions. From this it is obvious that the IR divergences in the heavy-light current originate from regions of phase space where either the gluon three-momentum $|{\bf k}|$ or the angle $\theta$ between the gluon and the light fermion goes to
zero. Other divergences arise if the three-momentum of the gluon goes to infinity or $\theta$ goes to $\pi$. These divergences are UV. To check if the IR divergences match between the two theories one has to use an IR regulator that regulates all IR divergences in both theories. To get insight into the behavior of the three-momentum and the
angle, it will be instructive to perform the required loop integrals
by integrating over $k_0$ using the method of residues, and then integrating
over the magnitude of the three-momentum and the solid angle. This will
allow us to identify clearly the IR divergences as described above. Let
us illustrate this method by showing that all $1/\epsilon$ divergences
in the \SCETa one-loop calculation of the previous section are UV. For
the first diagram we find
\begin{eqnarray}
A^{Ia}_{p_c} &=& -ig^2 C_F \mu^{4-d}\! \int \!\! \frac{{\rm d}^dk}{(2\pi)^d}  
\frac{1}{[\tilde p_c -n \!\cdot\! k+i 0][v \!\cdot\! (p_s-k)+i 0]
[k^2+i 0]} \nonumber\\
&=& -\frac{g^2 C_F}{2}\frac{\Omega_{d-2}}{(2\pi)^{d-1}} \mu^{4-d} \! \int_0^\infty
\!\!\!\!\!\! {\rm d}|{\bfk}||{\bfk}|^{d-2} \! \int_{-1}^1 \!\!\!\!\!
{\rm d}\!\cos\theta \, \sin^{d-4}\theta
\frac{1}{\left(|{\bfk}|(1-\cos\theta)-\tilde p_c\right)({|\bfk|}- v
\!\cdot\! p_s)|{\bfk}|}\, .
\end{eqnarray}
Performing the remaining integrals, we of course reproduce the result
obtained previously, but this form demonstrates that all divergences
from regions $|{\bfk}| \to 0$ and $(1-\cos \theta) \to 0$ are
regulated by the infrared regulators and thus all $1/\epsilon$
divergences are truly UV.

The second diagram is
\begin{eqnarray}
\label{AIIbintegrand}
A^{Ib}_{p_c} 
&=& -2 i g^2 C_F \mu^{4-d}\! \int \!\! \frac{{\rm d}^dk}{(2\pi)^d} 
\frac{\bn \!\cdot\! (p_c-k)}
{[-\bn \!\cdot\! k + i 0][(p_c-k)^2+i0][k^2+i0]}
\nonumber\\
&=& -2 i g^2 C_F \left[I_1+I_2\right]\, ,
\end{eqnarray}
where $I_1$ and $I_2$ are the integrals with the $\bn \!\cdot\! p$ and the
$\bn \!\cdot\! k$ terms in the numerator, respectively. The integral $I_2$
is standard and we find
\begin{eqnarray}
I_2 = \frac{i}{16 \pi^2} \left[\frac{1}{\epsilon} - \log
\frac{-p_c^2}{\mu^2} +2\right]\, ,
\end{eqnarray}
where $\epsilon$ regulates only UV divergences. 
For the first integral we again perform the $k_0$ integral by contours and we find
\begin{eqnarray}
\label{I1}
I_1 &=& \frac{i \bn \!\cdot\! p_c}{2} \frac{\Omega_{d-2}}{(2\pi)^{d-1}} \mu^{4-d} \! \int_0^\infty
\!\!\!\!\!\! {\rm d}|{\bfk}||{\bfk}|^{d-2} \! \int_{-1}^1 \!\!\!\!\!
{\rm d}\!\cos\theta \, \sin^{d-4}\theta 
\nonumber \\
&&\bigg[ - \frac{1}{{\bfk}^2 (1+\cos\theta)
[2|{\bfk}|(p_0-|{\bf p}|\cos\theta)-p_c^2]}
+ \frac{1}{a[p_0+a+|{\bfk}| \cos\theta][2p_0^2+2p_0a
-2|{\bfk}||{\bf p}| \cos\theta-p_c^2]} \bigg]\, ,
\end{eqnarray}
where
\begin{eqnarray}
p_c=(p_0,{\bf p})\, , \qquad a = \sqrt{\bfk^2 + {\bf p}^2
-2|\bfk||{\bf p}| \cos\theta}\, .
\end{eqnarray}
From this expression we can again see that all IR singularities from
$|{\bfk}| \to 0$ and $(1-\cos\theta) \to 0$ are regulated by the
off-shellness, and all remaining divergences are UV. Note furthermore
that in the limit $|{\bfk}| \to \infty$, with unrestricted $\theta$,
the two terms cancel each other, so that there is no usual UV
divergence. This agrees with the fact that there are five powers of
$k$ in the denominator of the integrand in
Eq.~(\ref{AIIbintegrand}). However, in the limit $|{\bfk}|\to \infty$
with $|{\bfk}| (1+\cos\theta) \to 0$ the second term of Eq.~(\ref{I1})
remains finite, whereas the first term develops a double
divergence. Thus, it is this region of phase space that gives rise to
the double pole in this diagram. The presence of the square roots
makes the evaluation of the remaining integrals difficult, but we have
checked that we reproduce the divergent terms of the result given in
Eq.~({\ref{AIIbresult}).

From the above discussion it follows that the off-shellness of the
external fermions regulates all the IR divergences, and that the
$1/\epsilon$ divergences all correspond to divergences of UV
origin. The situation is different in \SCETb, since the off-shellness
of the light quark does not enter diagram (a). We find
\begin{eqnarray}
A^{IIa}_{p_c} &=& -ig^2 C_F \mu^{4-d} \! \int \!\! \frac{{\rm d}^dk}{(2\pi)^d} 
\frac{1}{[-n \!\cdot\! k+i 0][v \!\cdot\! (p_s-k)+i 0][k^2+i 0]} 
\nonumber\\
&=& -\frac{g^2 C_F}{2} \frac{\Omega_{d-2}}{(2\pi)^{d-1}} \mu^{4-d} \! \int_0^\infty
\!\!\!\!\!\! {\rm d}|{\bfk}||{\bfk}|^{d-2} \! \int_{-1}^1 \!\!\!\!\!
{\rm d}\!\cos\theta \, \sin^{d-4}\theta  \frac{1}{{\bfk}^2(1-\cos\theta)
(|{\bfk}|-v \!\cdot\! p_s)}\, .
\end{eqnarray}
The IR divergence originating from the limit $(1-\cos\theta) \to 0$ is
not regulated by the off-shellness. Thus part of the $1/\epsilon$
divergences in Eq.~(\ref{AIIaneubertresult}) are of IR origin. In
other words, the fact that the terms logarithmic in the off-shellness
in the \SCETa amplitude Eq.~(\ref{A1psq}) are not reproducing the
corresponding terms in the \SCETb amplitude Eq.~(\ref{A2psqinterm}) does not
imply that the IR divergences do not match between the two
theories. In order to check whether the IR divergences of the two
theories match, one needs a regulator that regulates all IR
divergences in both \SCETa and \SCETb.

As an alternative IR regulator one could try to use a small gluon mass. Consider the first diagram in \SCETa again, this time
with a gluon mass. We find
\begin{eqnarray}
\label{AIagluonmass}
A^{Ia}_{m} &=& -ig^2 C_F \mu^{4-d} \! \int \!\! \frac{{\rm d}^dk}{(2\pi)^d} 
\frac{1}{[-n \!\cdot\! k+i 0][v \!\cdot\! (p_s-k) +i 0][k^2-m^2+i 0]} 
\nonumber\\
&=& -\frac{g^2 C_F}{2}\frac{\Omega_{d-2}}{(2\pi)^{d-1}} \mu^{4-d} \! \int_0^\infty \!\!\!\!\!\! {\rm d}|{\bfk}| |{\bfk}|^{d-2} \! \int_{-1}^1
\!\!\!\!\! {\rm d}\!\cos\theta \, \sin^{d-4}\theta
\frac{1}{(\bfk^2+m^2 - v \! \cdot \! p_s \sqrt{\bfk^2+m^2} )(\sqrt{\bfk^2+m^2}-|{\bfk}|\cos\theta)}\, .\,\,\,\,\,\,\,\,
\end{eqnarray}
Again, all divergences $|{\bfk}| \to 0$ and $(1-\cos\theta) \to 0$ are 
regulated by the gluon mass, but in the limit $|{\bfk}| \to \infty$ with 
$|{\bfk}|(1-\cos\theta)\to 0$ the integrand becomes
\begin{eqnarray}
\label{gluonproblem}
\frac{|\bfk|^{d-4} \sin^{d-4}\theta}{|{\bfk}|(1-\cos\theta) +
\frac{m^2}{2|{\bfk}|}}\, ,
\end{eqnarray}
so that the term that could potentially regulate the
$(1-\cos\theta)\to 0$ divergence goes to zero as $|\bfk| \to
\infty$. This is why a gluon mass cannot be used to regulate the IR of
SCET.

\subsection{A new regulator for SCET}

The gluon mass is not an appropriate IR regulator for SCET because it
appears in the combination $m^2/|{\bfk}|$ in the expression
(\ref{gluonproblem}). Instead of using a gluon mass, consider adding
the terms~\footnote{An alternative regulator has been introduced in
Ref.~\cite{analytic}. In that paper a quark mass is used in conjunction with an ``analytic'' regulator, which regulates the $(1-\cos\theta) \to 0$ divergence. The
conclusions about the soft-collinear mode in Ref.~\cite{messenger} are
similar to the ones drawn here. However, we believe that a regulator
such as the one introduced here is advantageous, since it can
naturally be defined at the level of the Lagrangian, and a single dimensionful parameter regulates all IR divergences. }
\begin{eqnarray} 
{\cal L}^c_{\rm reg} &=& - \frac{\delta}{2}A^c_\mu
\bnP A_c^\mu \nn\\
{\cal L}^{(u)s}_{\rm reg} &=& - \frac{\delta}{2}A^{(u)s}_\mu
i \bn \!\cdot\! \partial A_{(u)s}^\mu 
\end{eqnarray} 
to the collinear and (u)soft gluon
Lagrangians. Here, $\bnP$ is the label operator which picks out the large momentum label of the collinear gluon field. Both of these terms are generated if a similar term is added to the full QCD gluon action before constructing SCET. Note that these terms preserve the invariance of the theory under the field redefinitions given in Eq.~(\ref{Yredef}). 

The infinitesimal, dimensionful parameter $\delta$ suffices to regulate all IR divergences in SCET, unlike the gluon mass. Following the same steps as in
Eq.~(\ref{AIagluonmass}). We find
\begin{eqnarray}
\label{AIanewregulator}
A^{Ia}_\delta &=& -\frac{g^2 C_F}{2}\frac{\Omega_{d-2}}{(2\pi)^{d-1}} \mu^{4-d} \!
\int_0^\infty \!\!\!\!\!\! {\rm d}|{\bfk}| |{\bfk}|^{d-2} \! \int_{-1}^1
\!\!\!\!\! {\rm d}\!\cos\theta \, \sin^{d-4}\theta  \frac{8}{b(\delta + b)
(\delta + b - 2|\bfk| \cos \theta)}\, ,
\end{eqnarray}
where
\begin{eqnarray}
b = \sqrt{4\bfk^2 + \delta^2 + 4 |\bfk| \delta
\cos\theta}\, .
\end{eqnarray}
Obviously, the parameter $\delta$ regulates the divergences $|\bfk| \to 0$ 
and $(1-\cos\theta) \to 0$, just as the gluon mass did. Expanding around 
the limit $|\bfk| \to \infty$ with $|\bfk|(1-\cos\theta)\to 0$ the integrand 
becomes
\begin{eqnarray}
\frac{|\bfk|^{d-4} \sin^{d-4}\theta}{|\bfk|(1-\cos\theta) + \delta
}\, ,
\end{eqnarray}
and this IR region is therefore regulated as well. Even though $\delta$ is enough to regulate all IR divergences in SCET, we will keep the heavy quark off its mass-shell for later convenience. 

Performing the integrals using the method above is difficult. While performing the $k_0$ integration using the method of residues gives
insight into the divergence structure of the loop integrals, it is simpler to
perform the integrals using the variables $n\!\cdot\! k$ and
$\bar{n}\!\cdot\! k$ instead. The first diagram in \SCETa with this new
regulator is then given by
\begin{eqnarray}
A^{Ia}_\delta &=& -ig^2 C_F \mu^{4-d}\! \int \!\! \frac{{\rm d}^dk}{(2\pi)^d} 
\frac{1}{[-n \!\cdot\! k+i 0][v \!\cdot\! (p_s-k)+i 0][k^2-\delta \bn \!\cdot\! k+i 0]} 
\nonumber\\
&=& -\frac{g^2 C_F}{2\pi} (4\pi)^{1-d/2} \Gamma \left( 2-\frac{d}{2}\right) 
\mu^{4-d} \! \int^{\infty}_{\delta} \!\!\!\!\!\! {\rm d} n \!\cdot\! k  \, 
\, n \!\cdot\! k^{-1} \left( n \!\cdot\! k-\delta \right)^{d/2-2} (n\!\cdot \! k - 2 v \!\cdot \! p_s)^{d/2-2}
\nonumber\\
&=& \frac{\alpha_s C_F}{4\pi} \left
[-\frac{1}{\epsilon^2}+\frac{2}{\epsilon}\log
\frac{\delta}{\mu}-2\log^2\frac{\delta}{\mu}
+ 2\log \left(1-\frac{2v\!\cdot\! p_s}{\delta}\right)  \log \frac{2v\!\cdot\! p_s}{\delta}
+ 2 \, {\rm Li}_2\left(1-\frac{2v\!\cdot\! p_s}{\delta}\right) - \frac{3 \pi^2}{4}
\right]\, .
\end{eqnarray}
Similarly, it is possible to show that the parameter $\delta$ regulates all IR 
divergences in the second diagram, for which we find
\begin{eqnarray}
A^{Ib}_{\delta} &=& -2 i g^2 C_F \mu^{4-d}\! \int \!\! \frac{{\rm d}^dk}{(2\pi)^d}  
\frac{\bn \!\cdot\! (p_c-k)}{[-\bn \!\cdot\! k+i0]
[k^2 - 2 k \!\cdot\! p_c+i0][k^2-\delta \bn \!\cdot \! k+i0]}
\nonumber\\
&=& \frac{\alpha_s C_F}{4 \pi} \left
[ \frac{2}{\epsilon^2}+\frac{2}{\epsilon}-\frac{2}{\epsilon}
\log\frac{\delta \, \bn \!\cdot\! p_c}{\mu^2} 
+\log^2\frac{\delta \, \bn \!\cdot\! p_c}{\mu^2}
-2\log\frac{\delta \, \bn \!\cdot\! p_c}{\mu^2} + 4 - \frac{\pi^2}{6}
 \right]\, .
\end{eqnarray}
The mixed UV-IR divergences cancel in the sum of the two diagrams,
\begin{eqnarray}
\label{A1delta}
A^I_\delta &=& \frac{\alpha_s C_F}{4 \pi} \left[ \frac{1}{\epsilon^2} +
\frac{2}{\epsilon} \log \frac{\mu}{\bn \!\cdot\! p_c}  + \frac{5}{2\epsilon}
+ \log^2 \frac{\delta \bn \!\cdot\! p_c}{\mu^2}  - 2 \log^2  \frac{\delta}{\mu}
-\frac{3}{2} \log \frac{\delta \bn \!\cdot\! p_c}{\mu^2}- 2 \log  \frac{\delta-2 v \!\cdot\! p_s}{\mu}\right.\nonumber\\
&&\qquad \left.
+ 2\log \left(1-\frac{2v\!\cdot\! p_s}{\delta}\right)  \log \frac{2v\!\cdot\! p_s}{\delta}
+ 2 {\rm Li}_2\left(1-\frac{2v\!\cdot\! p_s}{\delta}\right) 
+ \frac{11}{2} - \frac{11 \pi^2}{12}
\right]\,,
\end{eqnarray}
and one can show that this result reproduces the IR behavior of full QCD.

Since the regulator is in the gluon action, it is the same for \SCETa
and \SCETb, and the two diagrams in \SCETb are identical to those in
\SCETa since the integrands are exactly equal:
\begin{eqnarray}
A^{II}_\delta = A^I_\delta\,.
\end{eqnarray}
Therefore, the IR divergences in
\SCETb are exactly the same as those in \SCETa. 

While in \SCETa it is possible to choose the scaling $\delta \sim Q\lambda^2$ such that both the contributions to the collinear and the usoft gluon action are leading order in the power counting, the same is not true in \SCETb. Choosing $\delta \sim Q\lambda^2$ to make the IR regulator leading order in collinear gluon Lagrangian  makes it suppressed by one power of $\lambda$ in the soft Lagrangian. This can be understood physically, since in going from \SCETa to \SCETb the typical scaling of the (u)soft momenta remains of order $\Lambda_{\rm QCD}$, while the off-shellness of the collinear particles is lowered. However, the IR divergence from $n\!\cdot \! k \to 0$ corresponds to $(1-\cos \theta) \to 0$, and the typical cutoff on $(1-\cos \theta)$ is set by the collinear scales. Since $n \!\cdot\! k_c \ll n \!\cdot\! k_s$ it is natural that any cutoff $\kappa$ regulating the $n \!\cdot\! k_s \to 0$ divergence will satisfy $\kappa \ll n \!\cdot\! k_s$.  This is not a problem, since the IR regulator does not introduce a new scale into the effective theory. 

If one insists on keeping the scaling manifest, one is forced to drop the regulator term in the soft gluon Lagrangian. In this case, the diagram (a) in \SCETb no longer includes the IR regulator $\delta$ and is therefore not regulated properly. The calculation then reduces to the result given in Eq.~(\ref{AIIaneubertresult}). Part of the $1/\epsilon$ divergences in this result are from true UV divergences, but others are due to the unregulated $(1-\cos \theta) \to 0$ IR divergences, which arise from physics at the scale $n\!\cdot\! k \sim Q\lambda^2$. These IR divergences can be recovered by adding a diagram containing a gluon scaling as $n \!\cdot\! k \sim Q \lambda^2$, $\bn \!\cdot\! k \sim Q\lambda$. Requiring $n \!\cdot\! k\, \bn \!\cdot\! k \, \sim k_\perp^2$, this is the soft collinear messenger mode introduced in \cite{messenger}. The resulting diagram (c) gives
\begin{eqnarray}
A^{IIc}_\delta &=& -2 ig^2 C_F \mu^{4-d}\! \int \!\! \frac{{\rm d}^dk}{(2\pi)^d}  
\frac{1}{[-n \!\cdot\! k+i 0][2 v \!\cdot\! p_s-\bn  \!\cdot\! k+i 0]
[k^2-\delta \bn \!\cdot\! k+i 0]} \nonumber\\
&=&  \frac{\alpha_s C_F}{4\pi} \left[ -\frac{2}{\epsilon^2} +
\frac{2}{\epsilon} \log \frac{-2 v \!\cdot\! p_s \, \delta}{\mu^2} 
- \log^2  \frac{-2 v \!\cdot\! p_s \, \delta}{\mu^2}
- \frac{\pi^2}{2}
 \right]\, .
\end{eqnarray}
Adding all the diagrams we find
\begin{eqnarray}
\label{A2delta}
&& \frac{\alpha_s C_F}{4 \pi} \left[ 
\frac{1}{\epsilon^2} +
\frac{2}{\epsilon} \log \frac{\mu}{\bn \!\cdot\! p_c}  + \frac{5}{2\epsilon}
+ \log^2 \frac{\delta \bn \!\cdot \! p_c}{\mu^2} 
-\frac{3}{2} \log \frac{\delta \bn \!\cdot \! p_c}{\mu^2}- 2 \log  \frac{-2 v \!\cdot\! p_s}{\mu}
\right.
\nonumber\\
&&\qquad \qquad \left.
+2\log^2 \left(\frac{-2 v \!\cdot p_s}{\mu}\right) - \log^2\left( \frac{-2 v\!\cdot p_s \delta}{\mu^2}\right)
+ \frac{11}{2} - \frac{\pi^2}{4}\right]\, ,
\end{eqnarray}
which again reproduces the \SCETa result for $\delta \ll v\!\cdot\! p_s$. From this discussion it follows that the presence of the soft collinear messenger mode depends on the precise implementation of the IR regulator in the theory. Since the definition of an effective theory should be independent of the regulator used for an explicit calculation,  one can view the soft-collinear messenger mode as part of the IR regulator.

The term added to the gluon Lagrangian breaks gauge
invariance.  However, in this regard it is on the same footing as a gluon mass.  Since the coupling of gluons to fermions is via a conserved
current, this breaking of gauge invariance is only a problem once
gluon self-interactions are taken into account. For the
renormalization of operators such as the heavy-light current
considered in this paper, this only arises at the two loop level. In
matching calculations, the IR divergences always cancel. Hence any IR
regulator, including the one proposed here is applicable to matching
calculations at any order in perturbation theory. More care has to be
taken when using this regulator to calculate operator mixing, and in
this case gauge non-invariant operators have to be included beyond one
loop order. The main advantage of the new regulator is that it
preserves invariance of \SCETa under the field redefinition given in
Eq.~(\ref{Yredef}).

\section{Conclusions}
We have considered the matching of the heavy-light current in \SCETa
onto the corresponding current in \SCETb, in particular addressing the
question whether all long distance physics in \SCETa is correctly
reproduced in \SCETb. Using the off-shellness of the external heavy
and light fermions, it was argued in
Refs.~\cite{messenger} that a new collinear-soft messenger
mode is required in \SCETb to reproduce all the long distance physics
of \SCETa. Regulating the IR in \SCETb with an off-shellness is problematic, since the off-shellness prevents performing the field redefinition required to decouple the usoft gluons from the collinear particles, which allows the matching onto \SCETb easily.  In this paper we investigated the relationship between IR regulators and the definition of \SCETb.  By performing the
$k_0$ loop integral by contours and then writing the remaining
integrals as ${\rm d}|\bfk| \, {\rm d}\cos \theta$, we showed explicitly
that the off-shellness leaves the IR angular divergence
$(1-\cos\theta) \to 0$ unregulated in \SCETb.

We then introduced a new regulator for SCET that regulates soft
($|\bfk| \to 0$) and collinear ($\cos \theta \to 1$) IR divergences
in both \SCETa and \SCETb. This regulator modifies the gluon action,
much like a gluon mass, and thus preserves the field redefinition required to decouple usoft gluons from collinear particles in SCET.  Using this regulator, we showed
explicitly that \SCETb as formulated in Refs.~\cite{BPS,scetgauge}
reproduces all the IR divergences of \SCETa. We also argued that any cutoff $\kappa$ regulating the collinear divergence has to satisfy $\kappa \ll n \!\cdot\! k_s$. Regulating \SCETb this way therefore naturally requires keeping a formally subleading regulator in the theory. 

We also showed that a soft-collinear messenger mode is required in the definition of the IR regulator if one insists on power counting the regulator in the same way as kinetic terms in the action. In this case, there are unregulated IR divergences left in soft diagrams, which are corrected by additional contributions from the soft-collinear mode. 

The new regulator introduced in this paper preserves the invariance of \SCETa under the field redefinitions (\ref{Yredef}), and is therefore useful in studying factorization theorems beyond tree level. 

\acknowledgments{We would like to thank Thomas Becher, Richard Hill, Bjorn Lange, Michael Luke, Matthias Neubert, David Politzer, Ira Rothstein, Iain Stewart
and Mark Wise for useful discussions. This work was supported by the
Department of Energy under the contract DE-FG03-92ER40701.}


\begin{thebibliography}{99}

\bibitem{BFL}
C.~W.~Bauer, S.~Fleming and M.~E.~Luke,
Phys.\ Rev.\ D {\bf 63}, 014006 (2001).


\bibitem{BFPS}
C.~W.~Bauer, S.~Fleming, D.~Pirjol and I.~W.~Stewart,
Phys.\ Rev.\ D {\bf 63}, 114020 (2001).

\bibitem{BS}
C.~W.~Bauer and I.~W.~Stewart,
Phys.\ Lett.\ B {\bf 516}, 134 (2001).

\bibitem{BPS}
C.~W.~Bauer, D.~Pirjol and I.~W.~Stewart,
Phys.\ Rev.\ D {\bf 65}, 054022 (2002).

\bibitem{ffprl}
C.~W.~Bauer, D.~Pirjol and I.~W.~Stewart,
Phys.\ Rev.\ D {\bf 67}, 071502 (2003)
[arXiv:hep-ph/0211069].

\bibitem{HQET}
A.~V.~Manohar and M.~B.~Wise,
Cambridge Monogr.\ Part.\ Phys.\ Nucl.\ Phys.\ Cosmol.\  {\bf 10}, 1 (2000).

\bibitem{scetgauge}
C.~W.~Bauer, D.~Pirjol and I.~W.~Stewart,
Phys.\ Rev.\ D {\bf 68}, 034021 (2003).

\bibitem{NeubertSCET2}
R.~J.~Hill and M.~Neubert,
Nucl.\ Phys.\ B {\bf 657}, 229 (2003)
[arXiv:hep-ph/0211018].

\bibitem{messenger}
T.~Becher, R.~J.~Hill and M.~Neubert,
arXiv:hep-ph/0308122;
T.~Becher, R.~J.~Hill, B.~O.~Lange and M.~Neubert,
arXiv:hep-ph/0309227.

\bibitem{Hillprivate}
R.~J.~Hill, private communication.

\bibitem{offshellGreen}
For a recent discussion on this point see
I.~Z.~Rothstein,
arXiv:hep-ph/0308266.

\bibitem{analytic}
M.~Beneke and T.~Feldmann,
arXiv:hep-ph/0311335.



\end{thebibliography}
\end{document}